\documentclass[prd,aps,showpacs,preprintnumbers,amssymb]{revtex4}
\usepackage{graphicx}
\usepackage{dcolumn}
\usepackage{bm}
\begin{document}
\title{
\hfill\vbox{\hbox{\small PP-04-024 \qquad\qquad}}\\
\hfill\vbox{\hbox{\small SU 4252-790 \qquad\qquad}}\\
Leptonic CP violation phases using an ansatz for the neutrino mass matrix
and
application to leptogenesis
}
 \author{Salah {\sc Nasri}}
 \email{snasri@physics.umd.edu}
 \affiliation{Department of Physics, University of Maryland, College Park, MD
 20742-4111,
USA}
 \author{Joseph {\sc Schechter}}\email{schechte@phy.syr.edu}
\affiliation{Department of Physics, Syracuse University,
Syracuse, NY 13244-1130, USA}
  \author{Sherif {\sc Moussa}}
 \email{sherif@asunet.shams.eun.eg}
 \affiliation{Department of Mathematics, Faculty of Science, Ain Shams
University,
Egypt}

\date{February 2004}

\begin{abstract}
We further study the previously proposed Ansatz, Tr$(M_\nu)$=0, for a
prediagonal light Majorana type neutrino mass matrix. If CP violation
 is neglected
this enables one to use the existing data on squared mass differences
to estimate (up to a discrete ambiguity) the neutrino masses themselves.
If it is assumed that only the conventional CP phase is present, the Ansatz
enables us to estimate this phase in addition to all three masses. If
it is assumed that only the two Majorana CP phases are present, the Ansatz
enables us to present a one parameter family of solutions for the
masses and phases. This enables us to obtain a simple ``global" view
of lepton number violation effects. Furthermore using an SO(10) motivation
for the Ansatz suggests an amusing toy (clone) model in which the heavy
neutrinos have the same mixing pattern and mass ratios as the light ones.
In this case only their overall mass scale is not known
(although it is constrained
by the initial motivation). Using this toy model we make a rough estimate
of the magnitude of the baryon to photon ratio induced by the
 leptogenesis mechanism.
Solutions close to the CP conserving cases seem to be favored.
\end{abstract}

\maketitle

\section{Introduction}
\label{one}

    Remarkably, the recent KamLAND \cite{kamland},
 SNO \cite{sno} and K2K \cite{k2k} experiments
have added so much to the results obtained from earlier
solar neutrino, atmospheric neutrino and accelerator
experiments \cite{earlierexpts} that our knowledge about the neutrino
masses
and presumed lepton mixing matrix is almost as great as
our knowledge of the corresponding quantities in the quark sector.
Still there is an uncertainty about the interpretation due to
the results of the LSND experiment \cite{lsnd}.
 However, this experiment will
be checked soon by the miniBoone collaboration so one can wait for
confirmation before considering whether there is really a problem
with the usual picture of three massive neutrinos. In any event, there
is a strong presumption that this knowledge will play an
important role in going beyond the standard model of electroweak
interactions.

    One detail is, of course, lacking compared to the quark case.
Since the neutrino oscillation experiments measure only
the differences of the neutrino squared masses, the neutrino masses
themselves are not known. According to the latest analysis \cite{mstv}
the best fit to these differences is:
\begin{eqnarray}
    m_2^2-m_1^2 &=& 6.9 \times 10^{-5} eV^2, \nonumber \\
    |m_3^2-m_2^2| &=& 2.6 \times 10^{-3} eV^2.
\label{massdifferences}
\end{eqnarray}

     Now, there is a simple {\it complementary}
 Ansatz for the 3 x 3 neutrino mass
matrix, $M_\nu$ which, with some assumptions, enables one
to obtain the neutrino masses themselves from
 Eq. (\ref{massdifferences}); it requires:
\begin{equation}
Tr(M_\nu) = 0.
\label{ansatz}
\end{equation}
  It should be remarked that $M_\nu$ is to be regarded as the
prediagonal neutrino mass matrix. Furthermore, in the
relation $m_1+m_2+m_3=0$ which evidently results if
the neutrino mass matrix is taken to be real symmetric, the individual
 masses may be either positive or negative. The negative masses
can be converted to positive ones by adding appropriate factors of $i$
in the diagonalizing matrix.

    Eq. (\ref{ansatz}) was motivated in \cite{bfns}
from the grand unified model, SO(10) \cite{so10} and in \cite{hz} by
noting
that it would hold if $M_\nu$ is the commutator of two other matrices,
as may occur in certain models. If CP violation is neglected there
are essentially two possible solutions of the Ansatz: either $m_1$ and
$m_2$
have the same sign and
are approximately equal to each other and to $-m_3/2$ or else
 $m_1$ and $m_2$ have the opposite sign and are approximately equal
to each other in magnitude but much larger than the magnitude of $m_3$.

    In the present paper we will take the point of view that the
Ansatz, Eq. (\ref{ansatz}), is motivated from SO(10). However,
 the analysis is of course not dependent on
 the motivation.
  The SO(10) motivation arises from the
observation that Eq. (\ref{ansatz}) is,
although it seems at first different,
 essentially the same
as the characteristic prediction of grand unification:
\begin{equation}
m_b = rm_{\tau}, \label{guteq}
\end{equation}
relating the mass of the b quark with the mass of the
 the tau lepton ($r\approx 3$ takes account of the running of masses
from the grand unification scale to the low energy hadronic scale
of about $1 \;GeV$ \cite{RGE}). Note that in SO$(10)$ the neutrino
mass matrix takes the form:
\begin{equation}
M_\nu =M_L - M_D^TM_H^{-1}M_D,
\label{seesaw}
\end{equation}
where $M_L$, $M_H$ and $M_D$ are respectively the mass matrices
of the light neutrinos, heavy neutrinos and heavy-light mixing
(or ``Dirac matrix"). To start with, $M_\nu$
is an arbitrary symmetric matrix. If it is real we have CP invariance.
 Generally the second,
seesaw \cite{secondterm} term is considered to
dominate. However, as explained in \cite{bfns}, the present model is
based on the assumption that the first term dominates. That might
 not be unreasonable since a rough order of magnitude estimate
for the second term would be $m_t^2/10^{17}$ or about $3 \times
10^{-4}$ eV. (The quantity $10^{17}$ includes a factor $r^2
\approx 10$).
 Thus the second term
could be negligible if neutrino masses are
 appreciably greater than this value.

    In \cite{bfns} the complementary Ansatz was mainly studied
for the case of real $M_\nu$. Here we will be primarily interested in
the more general complex case which allows for non zero CP violation.
Furthermore, the input squared mass differences were not taken
to be very similar to
those in Eq.(\ref{massdifferences}) but were based on a least squares
fit \cite{os} of many different experiments {\it including} LSND. Here we
shall
adopt the more conventional values given in Eq.(\ref{massdifferences}).
A related analysis of Eq. (\ref{ansatz}) was recently made
in \cite{j}.

    For an understanding of the interesting leptogenesis mechanism
\cite{leptogen} of
baryogenesis it is important to also study the properties of the heavy
neutrinos which appear there.
 In the present SO(10) motivated framework this task
 turns out to be
remarkably simple;
the heavy neutrino mass matrix is given by
\begin{equation}
M_H = cM_\nu ,
\label{clone}
\end{equation}
where c is a numerical constant. This means that the eigenvalues
of $M_H$, to be denoted as $M_1, M_2, M_3$ are simple multiples
of the light neutrino masses $m_1, m_2, m_3$. In addition the
unitary matrix, $U$ which brings $M_\nu$ to diagonal form via
\begin{equation}
U^TM_{\nu}U=diag(m_1,m_2,m_3)= {\hat M_\nu},
\label{diagonalize}
\end{equation}
also diagonalizes $M_H$. In other words the heavy neutrinos
are clones of the light neutrinos in this picture. The result follows
from the choice of Higgs fields in SO(10). Trilinear Yukawa terms
which supply fermion masses can contain Higgs fields in the 10, 120 and 126
dimensional representations. To get the result just mentioned we need
to require that there is only one ``126" representation present,
although any number of ``10"'s and ``120"'s are allowed. Of
course we are also assuming the second term in Eq. (\ref{seesaw})
to be negligible for the purpose of generating the light neutrino
masses.

    In section II, we give our conventions for the lepton
 mixing matrix, including one conventional and two Majorana
type CP violation phases. An approximate equation relating
the complementary Ansatz to the parameters of the mixing matrix
and the physical light neutrino masses is written down. The
solutions for the neutrino masses in the CP conserving case,
based on the results of neutrino oscillation experiments,
are reviewed. In section III, the Ansatz equation is solved on
the assumption that only the conventional CP phase, $\delta$
is non zero. It is found that the only solutions correspond
to maximal phase, $sin^2\delta=1$ and neutrino masses close
to the ones obtained in the CP conserving case. In section
IV we investigate the more complicated but very interesting case
when only the two Majorana CP violation phases are non zero. In
this case there is a family (modulo a discrete
ambiguity) of solutions. We choose the mass of the third
 light neutrino, $m_3$ as our free parameter and calculate
the remaining neutrino masses and the Majorana phases as functions
of $m_3$. The model gives a lower bound for $m_3$ and
 the cosmology criterion on the sum of neutrino masses effectively
yields an upper bound. The results for the full range are scanned
numerically and a simple analytic interpretation of the pattern
is presented. The neutrinoless double beta decay parameter,
$|m_{ee}|$ is also calculated for each value of $m_3$. In
section V we make a rough estimate of the baryon to photon
ratio based on the leptogenesis mechanism. In order to do this it is
necessary to make some statements on the masses and mixings
of the heavy neutrinos. Our motivation for the original
Ansatz suggests a ``clone" model in which the heavy neutrinos have the
same mass ratios and mixing matrix as the light ones. The
only new parameter besides $m_3$ is the overall mass scale,
which however is constrained
by the original motivation to be somewhat on the large side. Then it is
relatively easy to calculate the lepton asymmetry parameters,
$\epsilon_i$ for the heavy neutrino decays as
 functions of mainly $m_3$. We combine these quantities in
a semi-quantitative way with criteria from previous treatments of the
Boltzmann evolution equations for the decaying neutrinos.
It is found that that the most plausible scenarios for
leptogenesis involve small CP violating Majorana phases and
light neutrino masses close to the ones predicted for the CP
conserving cases..
Finally section VI contains a brief discussion and a brief summary.

\section{Relating the ansatz to experiment}

   Here we will obtain an approximate equation which will be useful
for relating the complementary ansatz to experimental
information on neutrino squared mass differences and mixing
angles in the general case where CP violation is allowed.
The notation is the same as in section III of \cite{bfns} which contains
more details. For convenience, we will use what seems to be
 the most common convention for the part of the leptonic
 mixing matrix, $K_{exp}$,
which is measured in the usual neutrino oscillation experiments. This
part can be constructed as a product of elementary  transformations
in the (12), (23) and (13) subspaces. For example in the (12) subspace
one has:
\begin{equation}
\omega_{12}(\theta_{12},\phi_{12})=\left[ \begin{array}{c c c}
cos \theta_{12}&e^{i\phi_{12}}sin \theta_{12}&0\\
-e^{-i\phi_{12}}sin \theta_{12}&cos \theta_{12}&0\\
0&0&1
\end{array} \right]
\label{onetwotransf}
\end{equation}
with clear generalization to the (23) and (13) transformations.

 Then the usual convention corresponds to the choice:
\begin{equation}
K_{exp}=\omega_{23}(\theta_{23},0)\omega_{13}(\theta_{13}, -\delta)\omega_{12}(\theta_{12},0),
\label{threeomegas}
\end{equation}
with three mixing angles and the CP violation phase $\delta$. Multiplying
out yields:
\begin{equation}
K_{exp}=\left[ \begin{array}{c c c}
c_{12}c_{13}&s_{12}c_{13}&s_{13}e^{-i\delta}\\
-s_{12}c_{23}-c_{12}s_{13}s_{23}e^{i\delta}&c_{12}c_{23}-s_{12}s_{13}s_{23}e^{i\delta}&c_{13}s_{23}\\
s_{12}s_{23}-c_{12}s_{13}c_{23}e^{i\delta}&-c_{12}s_{23}-s_{12}s_{13}c_{23}e^{i\delta}&c_{13}c_{23}\\
\end{array} \right]
\label{usualconvention}
\end{equation}

where $s_{ij} = sin \theta_{ij}\; and \; c_{ij} = cos
\theta_{ij}$. Since the neutrinos are of Majorana type in this
model, there are expected also to be physical CP violating effects
due to
 the Majorana phases \cite{{sv}, {bhp}, {dknot}, {svandgkm}}. These may be
introduced via a unimodular diagonal matrix of phases,
\begin{eqnarray}
\omega_0(\tau)&=&diag(e^{i\tau_1}, e^{i\tau_2}, e^{i\tau_3}), \nonumber \\
\tau_1 + \tau_2 + \tau_3 &=& 0.
\label{majoranaphases}
\end{eqnarray}

The full lepton mixing matrix is then expressed as,
\begin{equation}
K=K_{exp}\omega_0^{-1}(\tau),
\label{fullmixingmatrix}
\end{equation}
which has three mixing angles and three independent CP violating
 phases. We
shall use this form in what follows.
As an aside, though, we remark that the full matrix could also
 be written \cite{bfns} in
an unconventional, but more symmetrical, way as:
\begin{equation}
K=\omega_{23}(\theta_{23}, \phi_{23})\omega_{13}(\theta_{13}, \phi_{13})
\omega_{12}(\theta_{12}, \phi_{12}).
\label{symmpresent}
\end{equation}

    As a final preliminary we need the leptonic W interaction term :
\begin{eqnarray}
{\cal L}&=& \frac{ig}{\sqrt 2}W_{\mu}^-{\bar e_L}\gamma_{\mu}K\nu + H.c.,
\nonumber\\
K&=&\Omega^{\dagger}U,
\label{weakinteraction}
\end{eqnarray}
where U is defined in Eq.(\ref{diagonalize}) and $\Omega^{\dagger}$ is
a unitary matrix which is needed to diagonalize the charged
lepton mass matrix. At this point we shall make the common approximation
that $\Omega$ can be replaced by essentially the unit matrix. This
is certainly not perfect but it seems reasonable for a start. Then
$U$ may be replaced by $K$, for which some elements are already well known.
This enables us to present the ansatz in the form:
\begin{equation}
Tr({\hat M_\nu}K_{exp}^{-1}K_{exp}^*\omega_0(2\tau))=0,
\label{ansatzwithKexp}
\end{equation}
where Eqs. (\ref{diagonalize}) and (\ref{fullmixingmatrix}) were
used. With the parameterized mixing matrix of Eq. (\ref{usualconvention})
the ansatz reads:

\begin{eqnarray}
m_1e^{2i\tau_1} \left[ 1 -2i(c_{12}s_{13})^2sin{\delta}e^{-i\delta} \right]
 +\nonumber\\
m_2e^{2i\tau_2} \left[ 1 -2i(s_{12}s_{13})^2sin{\delta}e^{-i\delta} \right]
\nonumber +\\
m_3e^{2i\tau_3} \left[ 1 +2i(s_{13})^2sin{\delta}e^{i\delta} \right] = 0.
\label{paramansatz}
\end{eqnarray}
In this equation we can choose the diagonal masses $m_1, m_2, m_3$ to be
 real positive.
However it will be a little more convenient in the CP conserving
case to allow some of them to be negative as well as positive. We
shall, for definiteness, mainly use the following best fit values
\cite{mstv} for the mixing angles:

\begin{equation}
s_{12}^2 = 0.30,\quad  s_{23}^2 = 0.50,\quad  s_{13}^2 = 0.003.
\label{exptmixingangles}
\end{equation}

It should be remarked that the precise value of $s_{13}$
is not well known, in contrast to the other two.

    Equation (\ref{paramansatz}) contains three unknown masses
and three unknown CP phases. It can be written as two real equations and
 augmented by two equations for two neutrino mass squared differences.
Thus there are four equations for six unknowns. By assuming some special
simplifications we can make the analysis tractable.

 For orientation let
 us first review the case when the theory is CP conserving so that all the
 three independent CP phases vanish. Then we will have 3 equations for
3 unknowns. The ansatz now reads $m_1+m_2+m_3=0$. Define:
\begin{eqnarray}
A&=&(m_2)^2-(m_1)^2 \nonumber \\
B&=&(m_3)^2-(m_2)^2
\label{AandB}
\end{eqnarray}
It can be deduced \cite{dfm} from the experimental data
that A is positive while the sign of B is not yet known.
Their magnitudes are given in Eq. (\ref{massdifferences}).
Thus there are two separate cases to be considered. First consider
both A and B positive. Then solving as in \cite{bfns} gives
the type I solution:

\begin{eqnarray}
m_1=0.0291 \textrm{ eV},\quad  m_2=0.0302 \textrm{ eV},\quad
m_3=-0.0593 \textrm{ eV}. \label{typeone}
\end{eqnarray}
 Next consider the type II solution where
B is negative; it gives:

\begin{eqnarray}
m_1=0.0503 \textrm{ eV},\quad  m_2=-0.0510 \textrm{ eV},\quad
m_3=0.00068 \textrm{ eV}. \label{typetwo}
\end{eqnarray}

Here $m_1$ and $m_2$ are still almost degenerate but differ in sign. However
$m_3$ is now relatively small compared to the others.

\section{Conventional CP violation}

     A fully predictive simple case would correspond to keeping $\delta$
as the only CP violation phase. Then the real part of the ansatz equation,
(\ref{paramansatz}) reads

\begin{equation}
m_1+m_2+m_3 -2s_{13}^2sin^2\delta(c_{12}^2m_1+s_{12}^2m_2+m_3)=0,
\label{realpart}
\end{equation}

while the imaginary part yields
\begin{equation}
s_{13}^2sin2\delta(c_{12}^2m_1 + s_{12}^2m_2 - m_3) = 0.
\label{imagpart}
\end{equation}
Note that the $m_i$'s are being taken real here, although they will be allowed
 to be either positive or negative. A negative $m_i$ is not a source of CP
violation even though it corresponds to a Majorana phase $\tau_i$
of $\pi/2$ when the masses are taken positive
\cite{negmass}. Now Eqs.(\ref{AandB}), (\ref{realpart}) and (\ref{imagpart})
constitute 4 equations for the three $m_i$'s and $\delta$.

    However it turns out that, except for the special case when
$s_{13}^2sin2\delta = 0$, there is no consistent solution of this
set of four equations for four unknowns. To see this, first consider
solving simultaneously the three equations (\ref{AandB}) and
(\ref{imagpart}) when the special case does not hold. The numerical
solution is seen to require $B<0$ and is found to be (with $A>0$):
\begin{equation}
m_1=-0.0548 \textrm{ eV},\quad m_2=0.0554 \textrm{ eV},\quad
m_3=-0.0217 \textrm{ eV}. \label{solutiontry}
\end{equation}
We must now check to see if this is consistent with the remaining
Eq. (\ref{realpart}). That leads to the requirement:
\begin{equation}
s_{13}^2sin^2{\delta}=\frac{1}{4}(1+\frac{m_1+m_2}{m_3})\approx 0.25,
\label{require}
\end{equation}
which, given the numerical value of $s_{13}^2$ in Eq.
(\ref{exptmixingangles}) clearly leads to the contradiction
$sin^2{\delta} \approx 80$. This contradiction will persist
   even if the upper bound (about 0.044) rather than the best fit
for $s_{13}^2$ is used. The result is also not changed
 if the signs of all the $m_i$'s are reversed.

    Thus the only possibility for pure $\delta$ type CP violation in
the present scheme is the special case where $sin^2{\delta}=1$. Then
we must solve simultaneously the three equations consisting
of Eq. (\ref{realpart}) in which this substitution has been
made for $sin^2{\delta}$ as well as Eqs. (\ref{AandB}). This results
in the equation for, say $m_2$,
\begin{equation}
sign(m_1)(m_2^2-A)^{1/2}(1-2s_{13}^2c_{12}^2)+m_2(1-2s_{13}^2s_{12}^2)
+(B+m_2^2)^{1/2}(1-2s_{13}^2)=0,
\label{m2equation}
\end{equation}
where $sign(m_3)$ has been arbitrarily taken positive. Knowing $m_2$,
the other two masses may of course be obtained from Eqs. (\ref{AandB}).

    Taking, for definiteness, the mixing angle from
Eq. (\ref{exptmixingangles}), one finds essentially two  different
solutions. These are quite similar to the Type I and type II
solutions given above in the CP conserving case. The type I
solution, with $B>0$ is
\begin{equation}
m_1=-0.0289 \textrm{ eV},\quad m_2=-0.0301 \textrm{ eV}, \quad
m_3=0.0592 \textrm{ eV}. \label{deltaI}
\end{equation}
The type II solution, with $B<0$, reads
\begin{equation}
m_1=0.0503 \textrm{ eV},\quad m_2=-0.0510 \textrm{ eV},\quad
m_3=0.00081 \textrm{ eV}. \label{deltaII}
\end{equation}
The very close similarity between the CP conserving
solutions and the solutions with $sin^2{\delta}=1$ is understandable
due to the small value of $s_{13}^2$.

\section{CP violation due to Majorana type phases}

   Since, as we have just seen, there is only one particular
allowed value for the conventional CP phase, $\delta$
if it is considered as the only source of CP violation
in the present scheme, it is of great interest to investigate
the Majorana phases. Clearly,
 it seems sensible to study these phases with the simplification
of putting $\delta$ to zero. From Eqs. (\ref{onetwotransf})
 and (\ref{threeomegas})
it is seen that the same effect is accomplished by setting
$s_{13}=0$. Then the
ansatz equation (\ref{paramansatz}) takes the form
\begin{equation}
m_1e^{2i\tau_1}+m_2e^{2i\tau_2}+m_3e^{2i\tau_3}=0.
\label{majtypeviol}
\end{equation}
For our present case it is  convenient to take all
three $m_i$'s to be real and positive ( note that a phase angle
$\tau_i =\pi/2$ corresponds to what was taken as a shorthand to be
a negative value of $m_i$).
 Together, Eq. (\ref{majtypeviol})
and Eq. (\ref{AandB}) comprise four real equations for five
unknowns (three masses and two independent $\tau_i$'s). To proceed
we shall thus assume a value for $m_3$ so that we have four
equations for four unknowns. In addition there is the two fold
ambiguity due to the unknown sign of B.
 Finally we shall allow $m_3$ to vary
to obtain a global picture of the situation.

\begin{figure}[htbp]
\centering
{\includegraphics[width=11.89cm,height=10.72cm,clip=true]{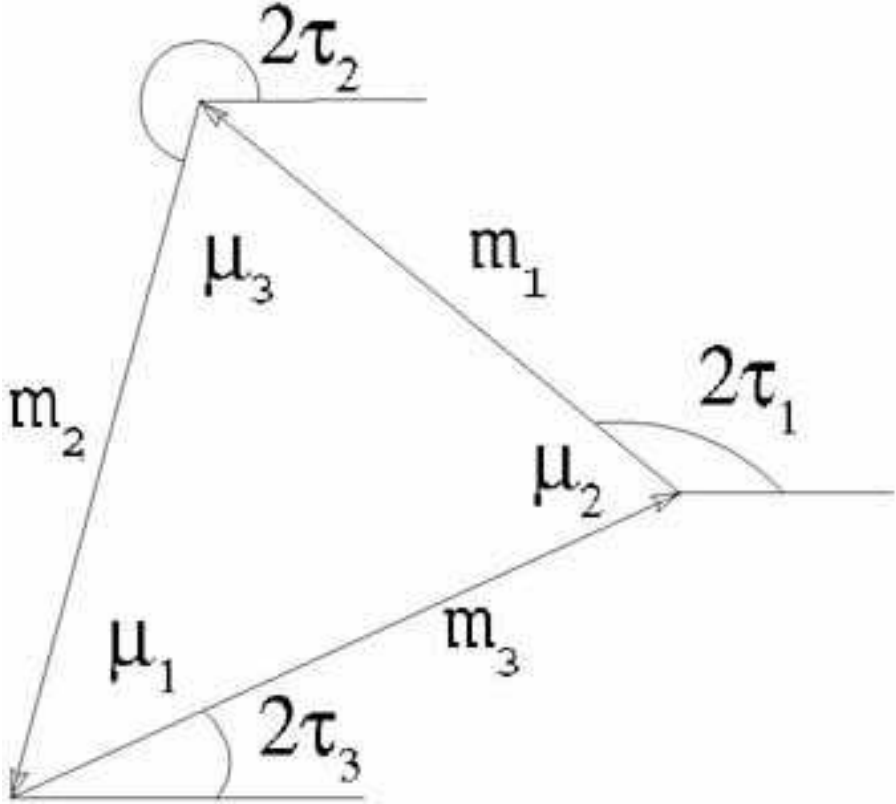}}
\caption[]{Vector triangle representing Eq. (\ref{majtypeviol}).}
\label{fig:1}
\end{figure}

   Now, once we have assumed a value for $m_3$, we can immediately find
$m_1$ and $m_2$ from Eqs. (\ref{AandB}). Furthermore, Eq. (\ref{majtypeviol})
    can be pictured in the complex plane
as a triangle formed from vectors with
lengths $m_i$, having angles $2\tau_i$ as measured from the
positive horizontal axis. Then let $\mu_i$ be the interior
angle opposite side $m_i$ as illustrated by the choice
of triangle in Fig. \ref{fig:1}.

 The problem reduces
to one from elementary plane geometry. Given the three sides ($m_i$),
of a triangle, find the three interior angles ($\mu_i$). We may start
for example, by using the law of
cosines to get
\begin{equation}
cos\mu_1=\frac{-m_1^2+m_2^2+m_3^2}{2m_2m_3},
\label{lawofcos}
\end{equation}
and continue similarly to get the others.
Finally the parameters $\tau_i$ which appear in the actual
 parameterization of Eq. (\ref{majoranaphases}) are found
from Fig. 1 as
\begin{eqnarray}
\tau_1&=& \frac{1}{6} (\pi-\mu_1-2\mu_2), \nonumber\\
\tau_2&=& \frac{1}{6} (\pi+2\mu_1+\mu_2), \nonumber\\
\tau_3&=& \frac{1}{6} (-2\pi-\mu_1+\mu_2).
\label{thethreetaus}
\end{eqnarray}
In particular, the quantities
\begin{eqnarray}
sin[2(\tau_1-\tau_2)]&=&-sin(\mu_1+\mu_2), \nonumber\\
sin[2(\tau_1-\tau_3)]&=&sin\mu_2, \nonumber\\
sin[2(\tau_2-\tau_3)]&=&-sin\mu_1,
\label{usefulsines}
\end{eqnarray}
will turn out to be of interest. Actually, given the three interior
angles $\mu_i$ of a triangle we do not get a unique choice of phase
differences $(\tau_i-\tau_j)$. While a rotation in the plane
of the triangle will not change these phase differences, it
is straightforward to see that the reflection of the triangle
about any line in the plane will reverse the signs of all the phase
differences. Thus there is another solution
in which an extra minus sign appears on each
right hand side of (\ref{usefulsines}).

    Now let us discuss the solutions of the complementary ansatz
equation for various assumed values of $m_3$.
In Table \ref{trianglesampler} the three real positive masses
as well as the corresponding values of the two independent
internal angles $\mu_1,\mu_2$ of the triangle are listed. Of course,
$\mu_1+\mu_2+\mu_3=\pi$.
The solution with $B>0$ (type I with
$m_3>m_2$) will be listed when it exists as well as the type II
solution ($B<0$ or $m_2>m_3$).

    Let us start with large values of $m_3$ and go down. Just from the
ansatz there is no upper bound on the value of $m_3$. However
there is a recent cosmology bound \cite{cosmobound} which
requires,
\begin{equation}
|m_1|+|m_2|+|m_3|<0.7 \textrm{ eV}. \label{cosmobound}
\end{equation}
Thus values of $m_3$ greater than about
0.3 eV are physically disfavored.
 Table \ref{trianglesampler} shows that at this value both
type I and type II solutions exist and correspond to
 almost equilateral
triangles. This is true also for higher values of $m_3$. Notice
that since the triangles are close to being equilateral, they have
large interior angles and hence ( see for example Eq.
(\ref{usefulsines})) large CP phases. The picture remains very
similar down to around $m_3=0.1$ \textrm{ eV} but as one gets
closer to the value,  roughly $0.0593$ \textrm{ eV},  where the
real type I solution
 of Eq. (\ref{typeone}) exists, there is a marked change.
It is seen that the interior angles of the type I solution
become small as it prepares to go to the degenerate triangle
corresponding to the real solution. We may get as small
CP phases as we like by tuning close to the real solution;
this is illustrated in Table \ref{trianglesampler}
 for a particular value
of $m_3$. If one further lowers $m_3$, it is found that
the type I solution no longer exists. On the other hand the type II
solution persists and does not change much until $m_3$
approaches the small value of roughly 0.00068 eV. There
are no solutions for $m_3$ smaller than this value. We can also tune
$m_3$ as illustrated in the table to get as small CP phases as
 we like for the type II case. It should be remarked that
the precise numbers in Table \ref{trianglesampler} are
based on the assumption that the best fit numbers given in
Eq. (\ref{massdifferences}) are exact and hence are meaningful
 to the accuracy given
only in the sense of comparing the various solutions with
 each other, not with experiment.
\begin{table}[htbp]
\begin{center}
\begin{tabular}{lllll}
\hline \hline type & $m_1,m_2,m_3$ in \textrm{ eV} & $\mu_1,\mu_2$
in radians & $|m_{ee}|$ in \textrm{ eV} &
  $\epsilon_1$, $\epsilon_2$, $\epsilon_3$ \\
\hline
\hline
I & 0.2955, 0.2956, 0.3000 & 1.038, 1.039 & 0.185 &  0.342, 0.433, 0.017
\\
II & 0.3042, 0.3043, 0.3000 & 1.055,1.056 & 0.187 & 0.330, 0.426, -0.0172
\\
I & 0.0856, 0.0860, 0.1000 & 0.946, 0.952 & 0.058 & 0.138, 0.060, 0.00137
\\
II & 0.1119, 0.1122, 0.1000 & 1.106, 1.111 & 0.065 & 0.194, 0.088-0.0024
\\
I & 0.0305, 0.0316, 0.0600 & 0.258, 0.268 & 0.030 &  0.00982, 0.00422,
0.00004 \\
II & 0.0783, 0.0787, 0.0600 & 1.172, 1.186 & 0.043 & 0.094, 0.041,-0.0011
\\
I & 0.0291, 0.0302, 0.0592715649 & 0.000552, 0.000574 & 0.030 & 1.96 $\times 10^{-6}$,
0.84 $\times 10^{-6}$, 0.71 $\times 10^{-7}$  \\
II & 0.0774, 0.0782, 0.0592715649 & 1.174, 1.188 & 0.042 &0.047, 0.020,
-0.0011  \\
II & 0.0643, 0.0648, 0.0400 & 1.243, 1.268 & 0.033 & 0.052,
0.023,-0.000681  \\
II & 0.0541, 0.0548, 0.0200 & 1.355. 1.417 & 0.024 & 0.018,
0.0078,-0.000335  \\
II & 0.0506, 0.0512, 0.0050 & 1.386, 1.658 & 0.021 &
 0.0057, 0.0025,-0.0000824  \\
II & 0.0503, 0.0510, 0.0010 & 0.814, 2.313 & 0.021 & 0.00073,
0.00031,-0.0000122  \\
II & 0.0503, 0.0510, 0.0006811 & 0.051361, 3.089536 & 0.021 &
 0.0000348, 0.0000150, -0.601 $\times 10^{-6}$  \\
\hline
\hline
\end{tabular}
\end{center}
\caption[]{Panorama of solutions as $m_3$ is lowered from about the
highest value which is experimentally reasonable to the
lowest value imposed by the model. In the type I solutions
 $m_3$ is the largest mass
while in the type II solutions $m_3$ is the smallest mass. For each
 value of $m_3$, the values of the model predictions for $m_1$
 and $m_2$ as well as the
 internal angles $\mu_1$ and $\mu_2$ are given. The model
prediction for the  neutrinoless double beta decay quantity $|m_{ee}|$
is next shown. Finally, the last column shows the estimated
 lepton asymmetries due to the decays of the heavy
 neutrinos. Note that the reversed sign
of lepton asymmetry is also possible, as discussed in the text.}
\label{trianglesampler}
\end{table}
    It is straightforward to give an analytic interpretation
of the pattern of solutions just observed. First note that CP
violation corresponds to a non degenerate triangle. Note also that the
orientation of the triangle in the complex plane is just obtained by
imposing the unimodularity condition for $\omega(\tau)$ in
Eq. (\ref{majoranaphases}). Hence the internal angles, $\mu_i$
are really the intrinsic carriers of CP violation. The determinant
of whether one has CP violation is the non vanishing of the quantity:
\begin{eqnarray}
{\cal A}&=&[m(m-m_1)(m-m_2)(m-m_3)]^{1/2}, \nonumber \\
m&=&\frac{1}{2} (m_1+m_2+m_3),
\label{arealaw}
\end{eqnarray}
which just expresses the area, ${\cal A}$
of a triangle in terms of the lengths of its sides.
This area may be rewritten in the convenient form:
\begin{equation}
{\cal A}=\frac{1}{4}\left([(m_1+m_2)^2-m_3^2][m_3^2-(m_1-m_2)^2]
\right)^{1/2}.
\label{secondarea}
\end{equation}
Now we may see that the vanishing of the first factor corresponds to
the type I real solution while the vanishing of the second factor
corresponds to the type II real solution. Furthermore, for a solution to
exist, the argument of the square root should be positive.
With the second factor, that establishes the minimum allowed value
of $m_3$ while with the first factor, that establishes the minimum
value of $m_3$ which allows a type I solution.

    An important test of the model is the experimental bound
 on neutrinoless
double beta decay. This implies \cite{ndbdexpt}
\begin{equation}
|m_{ee}| < (0.35 \rightarrow 1.30) \textrm{ eV}, \label{ndbd1}
\end{equation}
where
\begin{equation}
|m_{ee}| = |m_1(K_{exp11})^2e^{-2i\tau_1}+m_2(K_{exp12})^2e^{-2i\tau_2}
+m_3(K_{exp13})^2e^{-2i\tau_3}|.
\label{ndbd2}
\end{equation}
Using the parameterization of Eq. (\ref{usualconvention}) and
approximating $s_{13} = 0$ (which is reasonable in the present model
since $m_3$ is never much larger than $m_1$ or $m_2$), this can be
 written simply as:
\begin{equation}
|m_{ee}| = [(c_{12}^2m_1)^2+(s_{12}^2m_2)^2+
2m_1m_2c_{12}^2s_{12}^2cos(\mu_1+\mu_2)]^{1/2}.
\label{ndbd3}
\end{equation}
Here Eqs. (\ref{thethreetaus}) were also used. Reading $s_{12}^2$ from
(\ref{exptmixingangles}) then enables us to calculate $|m_{ee}|$ for
each line of Table I. It is seen that $|m_{ee}|$ decreases smoothly
with decreasing $m_3$ for each of the type I and type II solutions.
 All the values of $m_3$ listed are consistent with
the present bound. It is interesting that an improvement of the
experimental bound by an order of magnitude \cite{Future} would
provide a good test
 of the model.

\section{Estimate for leptogenesis}

    When one adopts the SO(10) motivation for the present
Ansatz, it turns out that the resulting model predicts
in a simple way the properties of the heavy neutrinos which
 are intrinsically contained in the SO(10) theory. This
feature may be used in connection with the leptogenesis
mechanism \cite{leptogen} of baryogenesis. According to this
 mechanism, the CP violating and lepton number violating
decays of the heavy neutrinos at a high
temperature (corresponding to
 the grand unification scale) in the very early universe
establish a lepton asymmetry. As the universe cools
further, the (B+L) violating but (B-L) conserving
 "sphaleron" interaction \cite{sphaleron} converts this into a
 baryon asymmetry which may be compared with the observed
ratio of baryons to photons in the universe. There are many
 interesting discussions of this mechanism in the literature
\cite{sakharov,reviews,lv,lpyetal,recent}.
Here, we will estimate the dependence on neutrino
masses and CP phases of the predicted baryon
asymmetry in the present model.

    The starting point of this discussion is the Yukawa
 term of the Lagrangian density which describes the tree level
 decay of a heavy Majorana neutrino, $N_j$ (where the
 subscript denotes a 3-valued generation index) to a
Higgs doublet member,
\begin{equation}
\Phi^c=\left(\begin{array}{c}{\bar \phi}^0 \\ -\phi^-\end{array}
\right)
\label{higgsdoublet}
\end{equation}
plus the appropriate member of the left-handed lepton doublet,
\begin{equation}
L_i=\left(\begin{array}{c}\nu_{iL} \\ e_{iL} \end{array}\right).
\label{leptondoublet}
\end{equation}
Then the Yukawa term reads:
\begin{equation}
{\cal L}_{Yukawa} =-\sum {\bar L}_i\lambda_{ij}\Phi^cN_j +
H.c.,
\label{yukawa}
\end{equation}
where $\lambda_{ij}$ is the matrix of Yukawa coupling constants.
We can simplify this expression, which is supposed to
contain the fermion fields in prediagonal bases, in several ways.
First, at the high temperature for which the N decays are relevant,
the phase transition to spontaneously broken SU(2) x U(1) has not
yet taken place. Thus we can consider the light fermions in $L_i$
to be massless and there is no need to insert suitable unitary matrices to bring
the light field mass matrices to diagonal form. However the heavy
neutrino, $N$ should be related to the physical field ${\hat N}$
with a unitary matrix $U$ as $N=U{\hat N}$. As mentioned in section I,
if the SO(10) model contains only a single "126" Higgs type field
(although any number of ``10"'s and ``120"'s are allowed)
and also if the first (non see-saw) term in Eq. (\ref{seesaw}) is dominant,
the prediagonal mass matrices for the light and heavy neutrinos must
be proportional to each other and the diagonalizing matrix $U$
must be the same one which appears in Eq. (\ref{weakinteraction}).
Approximating, as we did earlier, $\Omega$ to be essentially the unit matrix
we can set $U \approx K$. If the model of section IV is adopted,
for example, we can specify $K$, including
CP phases, to a fair approximation for each assumed
value of $m_3$. Finally we approximate the matrix of Yukawa couplings
by:
\begin{equation}
\lambda_{ij} \approx \frac{\delta_{ij}m_i^U}{r^{\prime}<\phi^0>},
\label{lambdaapprox}
\end{equation}
where $m_i^U$ are the three charge 2/3 quark masses at a low energy scale,
$r^{\prime} \approx 3$ is a suitable factor for running
these masses from the grand unified scale to the
low energy scale and $<\phi^0>\approx 246/{\sqrt 2}$
GeV. Note that ${\cal L}_{Yukawa}$ is the term responsible for generating
the neutrino Dirac matrix, $M_D$ in Eq. (\ref{seesaw}). In the simplest
approximation to the SO(10) theory the charge 2/3 quark mass matrix
and neutrino Dirac mass matrix are proportional to each other
and diagonal (since the quark mixings are after all small).

    Putting these things together we arrive at the ``effective"
term for calculating the heavy neutrino decays (at grand unified scale
 temperature):
\begin{equation}
{\cal L}_{Yukawa} = -\sum {\bar L}_ih_{ij}\Phi^c{\hat N}_j + H.c.,
\label{decayvertex}
\end{equation}
where
\begin{equation}
h_{ij} \approx \frac{m_i^UK_{expij}e^{-i\tau_j}}{<\phi^0>r^{\prime}}.
\label{hmatrix}
\end{equation}
The quantities needed for the calculation are the matrix
products $(h^{\dagger}h)_{ij}$. We may further simplify these
products by noting that the top quark mass is much heavier
 than the others so the products approximately become $h_{i3}^{\dagger}
h_{3j}$. Specifically, the diagonal products are:
\begin{eqnarray}
(h^{\dagger}h)_{11} &\approx& (s_{12}s_{23}/r^\prime)^2 ,
\nonumber \\
(h^{\dagger}h)_{22} &\approx& (c_{12}s_{23}/r^\prime)^2 ,
\nonumber \\
(h^{\dagger}h)_{33} &\approx& (c_{12}/r^\prime)^2 ,
\label{diagonalproducts}
\end{eqnarray}
where we used the numerical coincidence that $m_t=<\phi>$.
Furthermore, we set $s_{13}=0$ in agreement with the model of section IV
( See the parameterization of Eq. (\ref{usualconvention})).
Numerically, with Eq. (\ref{exptmixingangles}) and
$(r^{\prime})^2 \approx 10$ one obtains $(h^{\dagger}h)_{11}
\approx 1.50\times10^{-2}$, $(h^{\dagger}h)_{22}
\approx 3.50\times10^{-2}$ and $(h^{\dagger}h)_{33}
\approx 7.00\times10^{-2}$. In terms of these diagonal products,
the tree level widths of the heavy neutrinos are given by,
\begin{equation}
\Gamma_i = \frac{(h^{\dagger}h)_{ii}M_i}{8\pi},
\label{Nwidths}
\end{equation}
where $M_i$ is the mass of the i\emph{th} heavy neutrino.
The off diagonal products play an important role in
determining the lepton asymmetry. They are explicitly
given in the model of section IV as:
\begin{eqnarray}
(h^{\dagger}h)_{12}&\approx&-s_{12}c_{12}s_{23}^2e^{i(\tau_1-\tau_2)}
/(r^\prime)^2, \nonumber \\
(h^{\dagger}h)_{13}&\approx&s_{12}s_{23}c_{23}c_{13}e^{i(\tau_1-\tau_3)}
/(r^\prime)^2, \nonumber \\
(h^{\dagger}h)_{23}&\approx&-s_{23}c_{12}c_{23}c_{13}e^{i(\tau_2-\tau_3)}
/(r^\prime)^2, \nonumber \\
(h^{\dagger}h)_{ij}&=&(h^{\dagger}h)_{ji}^\ast,
\label{offdiagproducts}
\end{eqnarray}
where the CP phases $\tau_i$ depend on the choice of $m_3$ as
explained in section IV.
Numerically, one has $(h^{\dagger}h)_{12}\approx -2.29\times10^{-2}exp[i(\tau_1
-\tau_2)]$, $(h^{\dagger}h)_{13}\approx
2.74\times10^{-2}exp[i(\tau_1
-\tau_3)]$ and $(h^{\dagger}h)_{23}\approx
-4.18\times10^{-2}exp[i(\tau_2
-\tau_3)]$. In arriving at these estimates from
Eq. (\ref{exptmixingangles}) we arbitrarily took all the
signs of the trigonometric functions to be positive. This
will not lead to any ambiguity since, for the application of interest,
the off-diagonal products must be squared.

    The lepton asymmetry $\epsilon_i$, due to the decay of the i{\emph th}
heavy neutrino is defined as the ratio of decay widths:
\begin{equation}
\epsilon_i = \frac{\Gamma(N_i\rightarrow L+\Phi)-\Gamma(N_i \rightarrow
{\bar L}+{\bar \Phi})}{\Gamma(N_i\rightarrow L+\Phi)+\Gamma(N_i \rightarrow
{\bar L}+{\bar \Phi})}.
\label{defineepsilon}
\end{equation}
In this formula $L+\Phi$ stands for all pairs of the types $e^-_j +\phi^+$
and $\nu_j + {\bar \phi}^0$. This is an effect which violates C and CP
conservation, in agreement with the requirement of Sakharov \cite{sakharov}.
To get a non-zero value one must include the interference between the tree
diagram from Eq.(\ref{decayvertex}) and the one loop diagrams (of
both ``self-energy" and ``triangle" types). If the masses of the heavy neutrinos
are well separated  the result \cite{leptogen,pil} is:
\begin{eqnarray}
\epsilon_i&=&\frac{1}{8\pi}\sum_{j\ne
i}\frac{Im[(h^{\dagger}h)_{ij}(h^{\dagger}h)_{ij}]}
{(h^{\dagger}h)_{ii}}f(M_j^2/M_i^2), \nonumber \\
f(x)&=&\sqrt{x}\left[\frac{1}{1-x}+1 -(1+x)ln(1+1/x)\right].
\label{firstepsilonformula}
\end{eqnarray}
Note that the contribution to the lepton
asymmetry of the lightest heavy neutrino is
expected to be the most important one for the final
calculation of baryon asymmetry \cite{reviews}.

    Now let us make numerical estimates for the lepton
 asymmetries when  all CP violation is due to Majorana
 phases (section IV). From Eq. (\ref{clone}) we relate
 the heavy neutrino
masses to the light neutrino masses simply as:
\begin{equation}
M_i=cm_i,
\label{hlformula}
\end{equation}
where $c$ is a real, positive constant.
This equation has earlier been used \cite{jpr} for the study
of leptogenesis in the framework of a left-right symmetric model.
It should be noted that renormalization group effects \cite{rge}
will modify the exact proportionality of the light and heavy
neutrino masses as well as the equality of the corresponding
diagonalizing matrices. This should be taken into account
for a more accurate treatment.
 In the model of section IV
the third neutrino is typically somewhat further away in mass
from the other two, which are always relatively close.
For example, in the type II
situation, $m_3$ is the lightest of the light neutrino
masses so $M_3$ will be the lightest of the heavy neutrino masses
and the  contribution to the lepton asymmetry is  $\epsilon_3$. Using
Eqs. (\ref{firstepsilonformula}), (\ref{hlformula}),
(\ref{offdiagproducts}) and (\ref{usefulsines}) we
obtain:
\begin{equation}
\epsilon_3 \approx \left[-4.27f([m_1/m_3]^2)sin\mu_2
+9.94f([m_2/m_3]^2)sin\mu_1\right]\times10^{-4}.
\label{epsilon3}
\end{equation}
Notice that $c$ has canceled out in this formula and
 also cancels out in the determination of
the angles $\mu_i$. Thus the lepton asymmetry
given by this formula does not depend on the overall scale
of the heavy neutrino masses.

    In the type I case, the heavy neutrino spectrum
consists of two nearly degenerate lighter states, ($N_1$,
$N_2$) and a heavier state, $N_3$. For
the corresponding asymmetries $\epsilon_1$ and $\epsilon_2$,
 the diagrams involving self energy type
 corrections are enhanced since an internal heavy neutrino
 line will be close to its mass shell. The formulas \cite{pil}
thus, for greater accuracy, involve the decay widths and
we will approximate:
\begin{eqnarray}
\epsilon_1&=&\frac{Im[(h^{\dagger}h)_{12}(h^{\dagger}h)_{12}]}
{(h^{\dagger}h)_{11}(h^{\dagger}h)_{22}}
\frac{(M_1^2-M_2^2)M_1\Gamma_2}{(M_1^2-M_2^2)^2 +M_1^2\Gamma_{2}^2},
\nonumber \\
\epsilon_2&=&\frac{Im[(h^{\dagger}h)_{12}(h^{\dagger}h)_{12}]}
{(h^{\dagger}h)_{11}(h^{\dagger}h)_{22}}\frac{(M_1^2-M_2^2)M_2\Gamma_1}
{(M_1^2-M_2^2)^2 +M_2^2\Gamma_1^2}.
\label{twoepsilons}
\end{eqnarray}
Again, we may replace the heavy neutrino masses using
 Eq. (\ref{hlformula}) and note that the factor $c$ cancels out.
Inserting numbers, we obtain:
\begin{eqnarray}
\epsilon_1&\approx&-\frac{1.39\times10^{-3}sin(\mu_1+\mu_2)m_1m_2(m_1^2-m_2^2)}
{(m_1^2-m_2^2)^2+1.94\times 10^{-6}m_1^2m_2^2} \nonumber \\
\epsilon_2&\approx&-\frac{5.96\times 10^{-4}sin(\mu_1+\mu_2)m_1m_2(m_1^2-m_2^2)}
{(m_1^2-m_2^2)^2+3.56\times 10^{-7}m_1^2m_2^2}.
\label{twonumepsilons}
\end{eqnarray}
The values of all these asymmetries for the range of possibilities
 are listed in the last column of table I.

 Furthermore it must be noted that, owing to the non-uniqueness of
sign for all of Eqs. (\ref{usefulsines}), reversing the signs of all the
lepton asymmetries also yields a solution corresponding to our initial
Ansatz.

   Although the scale of the heavy neutrinos has been seen to cancel out
of the formulas (\ref{epsilon3}) and (\ref{twonumepsilons}) for the
lepton asymmetries in favor of their ratios (which are the same as those
of the light neutrinos in this model), there is nevertheless a consistency
condition implied by the SO(10) motivation for the starting Ansatz. This arises
since Eq. (\ref{yukawa}) is not only the source of the lepton asymmetry but
also provides the seesaw contribution to the light neutrino masses. For our
motivation we assumed that this contribution was dominated by the
first term of Eq. (\ref{seesaw}). To make a rough estimate of what this
means we assume all matrices of the seesaw term to be diagonal. Then
the value of $c^{1/2}$ defined in Eq. (\ref{hlformula}) should be greater
than $m_i^U/(m_ir^{\prime})$ in order that the first term
of Eq. (\ref{seesaw}) be greater than the second term. In the case of the type I
solution with $m_3\approx 0.06$ eV shown in Table I, this implies that the
lightest heavy neutrino should be heavier than about 2.6 $\times10^{13}$ GeV.
For the case of the type II solution with $m_3\approx7\times 10^{-4}$ eV, the
lightest heavy neutrino should be heavier than about 4.4 $\times10^{15}$ GeV.

    The goal of the baryogenesis problem is to understand the
ratio $\eta_B =n_B/n_{\gamma}$, the net baryon number density divided
by the photon density. Experimentally, this quantity is
 found\cite{cosmobound}, from the study of big-bang
 nucleosynthesis, to be
\begin{equation}
\eta_B = (6.5 \pm 0.4) \times 10^{-10}.
\label{eta}
\end{equation}
To obtain non-zero $\eta_B$,
it is not sufficient, as pointed out by Sakharov\cite{sakharov},
just to have non-zero values of the lepton asymmetry $\epsilon_i$, defined
in Eq. (\ref{defineepsilon}). In addition, the CP violating decays
of the heavy neutrinos must occur out of thermal
equilibrium. A detailed treatment requires solution of the Boltzmann
evolution equations for the system \cite{bdp}.
Here we shall make a rough estimate
which we use to draw
what might be a fairly robust conclusion.

    First, we should remark that the baryon asymmetry generated
by the sphaleron mechanism would be about -1/3 \cite{BL}, (for a
review see  \cite{reviews}) of an initial lepton asymmetry. The
lepton
 number violating decays of the i\emph{th} heavy neutrino
are usually roughly taken to be out
 of equilibrium if the decay rate $\Gamma_i$ in Eq. (\ref{Nwidths}) is
less than the Hubble rate,
\begin{equation}
H \approx 1.7g_*^{1/2}T^2/M_P ,
\label{hubble}
\end{equation}
where $g_*\approx 100$ is the number of effective light degrees of
freedom at the leptogenesis scale, $T$ is the temperature (
corresponding to the mass of the decaying heavy neutrino) and $M_P
\approx 1.22 \times 10^{19}$ \textrm{ GeV}. In the present model
this ratio takes the explicit form:
\begin{equation}
K_i=\frac{\Gamma_i}{H}\approx \frac{(h^{\dagger}h)_{ii}M_P}{427M_i},
\label{Kfactor}
\end{equation}
which is seen to be inversely proportional to $M_i$.
The net baryon asymmetry is estimated as \cite{reviews},
\begin{equation}
\eta_B \approx -\frac{7}{3g_*}\sum \epsilon_iD_i,
\label{etaformula}
\end{equation}
where the $D_i$ are suppression factors to be obtained
by numerical solution \cite{bdp} of the Boltzmann
equations. It is generally accepted that only the
contributions of the lightest heavy neutrinos
should not get washed out; thus we will set $D_i$ =0
for the heavier neutrinos.
   If $10 < K_i\lesssim 10^6$, the suppression factor
is often approximated by the analytic form \cite{kt}
\begin{equation}
D_i \approx \frac{0.3}{K_i[ln(K_i)]^{0.6}}.
\label{suppressionfactor}
\end{equation}
 When $K_i <1$, the suppression factor
is expected to be of order unity if $M_i$ is not too large.
However, as $M_i$ gets larger there is a sizeable washout effect
\cite{washout}.

    Glancing at the last column in table I and comparing with
the experimental value of $\eta_B$ in Eq. (\ref{eta}) as well as
Eqs. (\ref{etaformula}) and (\ref{suppressionfactor}) suggests that the
values of $\eta_B$ obtained for typical values of the assumed
light neutrino mass parameter $m_3$ would be considerably larger
than the experimental baryon asymmetry.
 However, we can expect to be able to
 obtain agreement with the experimental value since, as discussed
in section IV, we may make the Majorana CP phases as small as we like
by continuously tuning the independently chosen variable, $m_3$ so that the
triangle of mass vectors gets arbitrarily close to one of the two
degenerate straight line cases which causes ${\cal A}$ in
Eq.(\ref{secondarea}) to vanish. Thus the solutions of the model
which would be consistent with the observed baryon asymmetry
correspond to neutrino masses more or less close
the real cases of either Eq. (\ref{typeone}) or Eq. (\ref{typetwo}).

    The qualitative points: i. that in the present model the value
of the free parameter, $m_3$ can always be tuned to be arbitrarily
close to its values for the two real solutions (so that the CP violation
and hence leptogenesis strength becomes as small as desired) and
ii. that the characteristic lepton asymmetries, 
$\epsilon_i$ for values of $m_3$ away from these two real solutions
are rather large, comprise the main result of our discussion
of the application of the $Tr(M_\nu)=0$ Ansatz to the baryogenesis
problem. These points lead to the expectation that the physical
value of $m_3$ is likely to be close to one of the two values in
Eq. (\ref{typeone}) or Eq. (\ref{typetwo}) and that this conclusion might
persist even when our simplifications are not made. A more accurate
treatment would include the features: a) effect of non-trivial 
charged lepton mixing matrix, b) renormalization group
 induced deviations from the ``clone" treatment of the
heavy Majorana neutrinos and c) full integration of the Boltzmann
evolution equations. Even though our main conclusion is a qualitative one,
it seems nevertheless an interesting exercise to find what
values of light and heavy neutrino masses corrrespond to the correct order 
of magnitude of the observed baryon asymmetry. Note that
the seeming great accuracy of the entries in Table I is not
meant for precise comparison with experiment, but for comparison
of the results of different $m_3$ choices with each other.

    Specifically, consider the tuned type I solution in table I with
$m_3 \approx$ 0.05927 eV. We noted
in the discussion after Eq. (\ref{twonumepsilons}) that
 this would correspond to heavy neutrino
``clone" masses  $(M_1, M_2, M_3)$ greater than about $(2.60, 2.70, 5.27)\times
10^{13}$ GeV, respectively. We assume that the two lighter neutrinos are
the important ones and set $D_3$ = 0.
 The ratios $(K_1, K_2)$ defined in
Eq. (\ref{Kfactor}) would then be less than about (16.5, 37.1)
and would result in suppression factors
$(D_1, D_2)$ greater than (0.010, 0.0037).
Using Eq. (\ref{etaformula}) and table I for $\epsilon_1$
and $\epsilon_2$ then gives
 $|\eta_B|$ = 5.4 $\times 10^{-10}$, close to
 the experimental value in Eq.(\ref{eta}). This can be adjusted by
further tuning $m_3$ or to some extent by varying the overall
mass scale of $(M_1,M_2,M_3)$.

    For the type II case, first consider the
solution in table I with $m_3\approx$ 0.0007 eV. As discussed before,
this would correspond to heavy neutrino ``clone" masses
$(M_1,M_2,M_3)$ greater than about (320, 320, 4.4) $\times 10^{15}$ GeV.
In this case, $M_3$ is the lightest of the three heavy neutrinos
and is assumed to be the relevant one. We thus set
$D_1=D_2=0$.
The ratio $ K_3$ given in Eq. (\ref{Kfactor})
is then about 0.45 and indicates that
the lightest heavy neutrino is, as desired, decaying
out of equilibium. However, because its mass
is considerably higher than that of the type I case just
discussed, there is more wash out \cite{washout},
$D_3 \approx 3.5 \times 10^{-5}$. Reading $\epsilon_3$ from
table I then gives $\eta_B \approx 5 \times 10^{-13}$, about
three orders of magnitude too small. Thus we must raise
the value of $m_3$ a bit. Backing off a little to the case
$m_3\approx$ 0.005 eV in table I increases the value of
$\epsilon_3$ and also allows us (in line with the dominance
of the first term in Eq.(\ref{seesaw})) to choose the lower
bound of $M_3$ to be  smaller, around 6 $\times 10^{14}$ GeV.
This results in an estimate $|\eta_B| \approx 16 \times 10^{-10}$,
which is the correct order of magnitude. One might wonder
whether the contributions to $\eta_B$ from $\epsilon_1$
and $\epsilon_2$ are completely washed out in a case like the present.
However, even if they were dominant, it would just require us
to tune more closely toward small $m_3$.

   Thus, if the model of CP violation with just the Majorana phases is correct,
the magnitude of the baryon to photon ratio can be understood when either
 the sum of the three light neutrino masses is about 0.118 eV
 and $|m_{ee}|$ =0.030 eV (type I) or the sum of the three light neutrino
masses is about 0.107 eV and $|m_{ee}|$= 0.021 eV (type II). In both cases
the CP violating Majorana phases are extremely small. That might suggest
a possible model in which a small CP violating perturbation due to some
separate  effect modifies an otherwise CP conserving lepton sector.

   We can also calculate the baryon to photon ratio in the model
of section III, where $\delta$ is the only CP violating phase. There
we noted that the only possible choices of $\delta$ consistent
with our Ansatz satisfy $sin^2\delta=1$. Then we have the type I
solution for light neutrino masses given in Eq. (\ref{deltaI}) and the
type II solution given in Eq. (\ref{deltaII}). The corresponding
CP violation factors are now (to first order in the small
parameter $s_{13}$) for the type I case:
\begin{eqnarray}
Im[(h^{\dagger}h)_{12}(h^{\dagger}h)_{12}]&\approx&\frac{s_{13}sin\delta
sin(2\theta_{12})}{(r^\prime)^4}c_{23}s_{23}^3, \nonumber \\
Im[(h^{\dagger}h)_{13}(h^{\dagger}h)_{13}]&\approx&-\frac{s_{13}sin\delta
sin(2\theta_{12})}{(r^\prime)^4}s_{23}c_{23}^3, \nonumber \\
Im[(h^{\dagger}h)_{23}(h^{\dagger}h)_{23}]&\approx&-Im[(h^{\dagger}h)_{13}
(h^{\dagger}h)_{13}],
\label{cpviolfactorI}
\end{eqnarray}
and for the type II case:
\begin{eqnarray}
Im[(h^{\dagger}h)_{12}(h^{\dagger}h)_{12}]&\approx&-\frac{s_{13}sin\delta
sin(2\theta_{12})}{(r^\prime)^4}c_{23}s_{23}^3, \nonumber \\
Im[(h^{\dagger}h)_{13}(h^{\dagger}h)_{13}]&\approx&\frac{s_{13}sin\delta
sin(2\theta_{12})}{(r^\prime)^4}s_{23}c_{23}^3, \nonumber \\
Im[(h^{\dagger}h)_{23}(h^{\dagger}h)_{23}]&\approx&Im[(h^{\dagger}h)_{13}
(h^{\dagger}h)_{13}],
\label{cpviolfactorII}
\end{eqnarray}
As in the cases where only the Majorana phases contribute to the CP violation,
the predicted lepton asymmetries, $\epsilon_i$ will typically lead to a value
of the baryon to photon ratio much larger
than the experimental one. In the present case it is not possible to  fine tune
$\delta$. The only possibility would be to fine tune $s_{13}$ to an extremely small
value. This seems more artificial since $s_{13}$ is not required to vanish
 in the CP conserving situation. In any event, an experimental
measurement of non-zero $s_{13}$ would, practically speaking, rule out this
case as a candidate for leptogenesis.

\section{Discussion and summary}

    In this paper, we investigated an Ansatz which
correlates
information about the four quantities in the light neutrino sector
which are not yet known from experiment; namely, the absolute
mass of any particular neutrino, the ``conventional" CP
violation phase and the two Majorana phases. Of course, with
 input from analyses of neutrino oscillation experiments, the
masses of the other two neutrinos can be found, up to a discrete
ambiguity, if the mass of one is specified. The results of the
present paper
can be used for calculating many quantities of experimental
interest like the neutrinoless double beta decay amplitude
factor $m_{ee}$ (presented in section IV) and various lepton number
violating decays.

    The Ansatz is not completely predictive, unless some assumptions
are made. We first reviewed the case of assumed CP conservation
(where just the three neutrino masses are obtained).Then we showed that if
only the ``conventional" CP violation phase is assumed
to be non-zero, its value is fixed by the Ansatz to be maximal. A
possibly  more interesting case appears if we assume that only the
two Majorana phases are non-zero. This enables us to scan the
limited allowed range of assumed neutrino mass, $m_3$ (say) and
find the other two neutrino masses as well as the two Majorana
phases for each value of $m_3$. The result seems to cut through
a ``cross section" of interesting possibilities which
are described in a simple way.
 The still more complicated case
without setting any of the three CP phases to zero gives a
two parameter family of solutions and will be treated elsewhere.
Another (common) assumption we made for a first analysis is that the
measured lepton
mixing matrix is dominated by the neutrino factor. This is consistent
with the finding in recent years that the mixing
in the neutrino sector is apparently much larger than
the mixing in the quark sector (which in models is usually
relatively small and similar to that of the charged lepton
factor).

    It seems relevant to discuss briefly the status of the motivations
for the complementary Ansatz we are using. One motivation, based
on a loop mechanism for generating neutrino masses was discussed
recently by He and Zee \cite{hz}. Our motivation \cite{bfns}
was based on the grand unification group SO(10). This group
is well known to have the elegant feature that it accomodates
one generation of elementary fermions as well as an extra (now
desired) neutrino field in its fundamental spinor irreducible
representation. Naturally, there are many possibilities
for doing a detailed calculation using this group. One may ask
whether it should be regarded as being derived from a superstring
theory, whether it should be supersymmetric, whether the
symmetry breakdown should be dynamical, whether the symmetry
breakdown should be induced by Higgs fields and if so what
kind and how many, etc? We focus, in our motivation, on the
conventional possibility of using Higgs fields since it seems
almost kinematical now (although since no Higgs field has yet
been seen one should keep an open mind).
Of course, there have been many interesting treatments
along these lines \cite{so10studies}. Our
Ansatz is suggested by a relation involving only the neutrino mass matrix
which might be true (or at least approximately correct)
in a large number of models. In SO(10), tree level masses from
 a renormalizable Lagrangian can be obtained by using any number of
Higgs mesons belonging to the 10, 120 and 126 dimensional
representations. However, examining the form of
 the predicted mass matrices shows that the following
fairly general relation \cite{jrs} holds:
\begin{equation}
Tr(M^D-rM^E) \propto Tr(M_L),
\label{gutformula}
\end{equation}
for any number of 10's and 120's but only a single 126 present.
Here $M^D$ and $M^E$ are respectively the prediagonal mass
matrices of the charge -1/3 quarks and charge -1 leptons
while $r\approx3$, as previously mentioned. $M_L$, which arises
from the 126 Higgs field Yukawa couplings, is the non seesaw
part of the light neutrino
mass matrix which appears in Eq. (\ref{seesaw}). Taking
traces cancels the contributions (antisymmetric matrices) of
 any 120 Higgs multiplets to the left hand side.
Then, assuming the transformations which bring $M^D$ and $M^E$
to diagonal form to be roughly close to the identity we observe that
the left hand side is approximately equal to $m_b-rm_{\tau}$,
which is about zero. In fact this is a characteristic prediction of
grand unification.
 In turn the right hand side gives us the starting Ansatz
when it is assumed that the non seesaw term dominates in Eq. (\ref{seesaw}).
Of course, if this domination is to hold the masses of the heavy neutrinos
should not be too low. The present paper is in effect exploring the range
of possibilities which exist when these assumptions are made in SO(10)
models. An interesting question is whether this kind of limit
or the pure seesaw limit gives a better description of nature, even
if both terms are actually required.

    We remark that SO(10) also gives another similar relation,
\begin{equation}
Tr(M^U-r^{\prime}M_D) \propto Tr(M_L),
\label{otherrelation}
\end{equation}
when only one 126 Higgs field exists. Here $M^U$ and $M_D$
respectively denote the prediagonal charge 2/3 quark mass matrix
and the prediagonal neutrino ``Dirac" matrix connecting the heavy
and light neutrino fields.

    An intriguing way to learn more about CP violation in the lepton
sector is the study of the leptogenesis mechanism of baryogenesis.
We saw that the treatment of this process simplifies when one adopts the present
SO(10) motivation. Then the light neutrino mass matrix $M_\nu$ and $M_L$
are approximately equal and proportional (due to the assumption of only one 126
field in the theory) to the heavy neutrino mass matrix $M_H$. The only
free parameter for the heavy neutrinos is their overall
mass scale and this should not be too small to preserve non seesaw
dominance. We showed in section V that it is easy to estimate the
lepton asymmetry parameters $\epsilon_i$ for a ``panorama" of values
of the independent variable $m_3$ since they are
actually independent of the overall heavy neutrino mass scale.
As far as the resulting baryon to photon ratio, $\eta_B$ (parameterized
in Eq. (\ref{etaformula})) is concerned, the typical values of the $\epsilon_i$
give $\eta_B$ much greater than the experimental one for
suppression factors $D_i$ of order unity. We observed that if the
suppression factors are not too small one can therefore always choose a value
of $m_3$ close  enough to one of the two essentially different
 CP conserving solutions so that the
Majorana phases are small enough to get experimental agreement for
$\eta_B$.
Using  estimates of the suppression factors taken from other earlier studies,
 we noted that  this
conclusion seems reasonable. Of course the study of the suppression factors
by solving the Boltzmann evolution equations is an important topic which involves
many subtleties and would repay further work in the present model.
Finally, the posible indication of very small CP phases
 might suggest a model in which the CP violation
in the lepton sector has a separate identifiable source.

\section*{Acknowledgments}
\vskip -.5cm It is a pleasure to thank Deirdre Black, Amir
Fariborz, Cosmin Macesanu , Mark Trodden and David Schechter
 for their kind help with various aspects of this work. We would like
to thank Samina Masood for checking part of Table I.
 The work of S.N is supported by National Science foundation grant No. PHY-0099544.
The work of J.S. is supported in part by the U. S. DOE under
Contract no. DE-FG-02-85ER 40231.


\begin{thebibliography}{9}

\bibitem{kamland}KamLAND collaboration, K. Eguchi et al, Phys. Rev.
Lett. {\bf 90}, 021802 (2003).

\bibitem{sno}SNO collaboration, Q. R. Ahmad et al, arXiv:nucl-ex/
0309004.

\bibitem{k2k}K2K collaboration, M. H. Ahn et al, Phys. Rev. Lett. {\bf
90}, 041801 (2003).

\bibitem{earlierexpts}For recent reviews see, for examples,
S. Pakvasa and J. W. F. Valle, arXiv:hep-ph/0301061 and
V. Barger, D. Marfatia and K. Whisnant, arXiv:hep-ph/0308123.

\bibitem{lsnd}LSND collaboration, C. Athanassopoulos et al, Phys. Rev.
Lett. {\bf 81}, 1774 (1998).

\bibitem{mstv} M. Maltoni, T. Schwetz, M. A. Tortola and J. W. F. Valle,
arXiv:hep-ph/0309130.

\bibitem{bfns}D. Black, A. H. Fariborz, S. Nasri and J. Schechter,
Phys. Rev. {\bf D62}, 073015 (2000).

\bibitem{so10}H. Georgi, {\it Particles and Fields} (1974), edited by
 C. E. Carlson (AIP, New York, 1975), p. 575; H. Fritzsch
and P. Minkowski, Ann. Phys. (NY) {\bf 93}, 193 (1975); Nucl. Phys.
{\bf B103}, 61 (1976).

\bibitem{hz}X.-G. He and A. Zee, Phys. Rev. {\bf D68}, 037302, (2003).

\bibitem{RGE}M. Chanowitz, J. Ellis and M. K. Gaillard, Nucl.
Phys. {\bf  B128}, 506 {1977}; A. J. Buras, J. Ellis, M. K.
Gaillard and D. V. Nanopoulos, {\it ibid} {\bf B135}, 66 {1978}.

\bibitem{secondterm}T. Yanagida. in Proceedings of the workshop on
unified theory and baryon number in the universe, edited by O. Sawada
 and A. Sugamoto, KEK report 79-18, Tsukuba, 1979, p. 95; M. Gell-Mann,
P. Ramond and R. Slansky, in Supergravity, edited by P. van Niewenhuizen
and D. Z. Freedmann (North-Holland, Amsterdam,1979); R. N. Mohapatra
and G. Senjanovic, Phys. Rev. Lett. {\bf 44}, 912 (1980).


\bibitem{os}T. Ohlsson and H. Snellman, Phys. Rev. {\bf D60}, 093007
(1999). See also R. P. Thun and S. Mckee, Phys. Lett. {\bf B439}, 123
(1998); G. Barenboim and F. Scheck, {\it ibid} {\bf B440}, 332 (1998);
 G. Conforto, M. Barrone and G. Grimani, {\it ibid}, {\bf B447}, 122
(1999); I. Stancu, Mod. Phys. Lett. {\bf A14}, 689 (1999); A. Acker
and S. Pakvasa, Phys. Lett. {\bf B397}, 209 (1997). Criticism of
this approach has been expressed by G. L. Fogli, E. Lisi, A. Marrone
and G. Scioscia, hep-ph/9906450.

\bibitem{j}W. Rodejohann, Phys. Lett. {\bf B579}, 127 (2004).

\bibitem{leptogen}M. Fukugita and T. Yanagida, Phys. Lett.
{\bf B174}, 45 (1986).

\bibitem{sv}J. Schechter and J. W. F. Valle, Phys. Rev. D {\bf 22}, 2227 
(1980).

\bibitem{bhp}S. M. Bilenky, J. Hosek and S. T. Petcov, Phys. Lett. {\bf 94B}, 495
(1980).

\bibitem{dknot}M. Doi, T. Kotani, H. Nishiura, K. Okuda and E. Takasugi, Phys. Lett.
{\bf 102B}, 323 (1981).

\bibitem{svandgkm}J. Schechter and J. W. F. Valle, Phys. Rev. {\bf D23}, 1666
(1981); A. de Gouvea, B. Kayser and R. N. Mohapatra, Phys. Rev.
{\bf D67}, 053004 (2002).

\bibitem{dfm}A. de Gouvea, A. Friedland and H. Murayama, Phys. Lett.
{\bf B490}, 125 (2000).

\bibitem{negmass}See the discussion in section IV of \cite{bfns} above and
L. Wolfenstein, Phys. Lett. {\bf 107B}, 77 (1981).

\bibitem{cosmobound}D. N. Spergel et al, arXiv:astro-ph/0302209;
S. Hannestad, arXiv:astro-ph/0303076.

\bibitem{ndbdexpt}H. V. Klapdor-Kleingrothaus et al, Eur. Phys. J. {\bf A12}
147 (2001).

\bibitem{Future} G. Gratta, Talk given at XXI International
Syposium on Lepton and Photon Interactions at High Energies, 11-16
August 2003, Fermi National Accelerator Laboratory, Batavia,
Illinois USA, IJMPA {\bf 19 A8},1115 (2004).

\bibitem{sphaleron}V. A. Kuzmin, V. A. Rubakov and M. E. Shaposhnikov,
 Phys. Lett. {\bf B155}, 36(1985).

\bibitem{sakharov} The original mechanism is given in A. D. Sakharov,
Pis'ma Zh. Eksp. Teor. Fiz. {\bf 5}, 24 (1967) (JETP Lett. {\bf 5}, 24 (1967)).

\bibitem{BL} J. A. Harvey and M. Turner, Phys. Rev. {\bf D42},
3344 (1990).

\bibitem{reviews}Reviews are given in E. W. Kolb and M. S. Turner, The Early
 Universe, (Addison-Wesley, Reading, MA, 1989); A. Pilaftsis, Int. J. Mod. Phys.
{\bf A14}, 1811 (1999)
; A. Riotto and M. Trodden, Annu. Rev. Nucl. Part. Sci. {\bf 45}, 35
(1999).

\bibitem{lv}M. A. Luty, Phys. Rev. {\bf D45}, 455 (1992); C. E. Vayonakis, Phys.
Lett. {\bf B286}, 92 (1992).

\bibitem{lpyetal}P. Langacker, R. D. Peccei and T. Yanagida, Mod. Phys. Lett.
{\bf A1}, 541 (1986); R. N. Mohapatra and X. Zhang, Phys. Rev. {\bf D45}, 5331 (1992);
K. Enqvist and I. Vilja, Phys. Lett. {\bf B299}, 281 (1993); H. Murayama, H. Suzuki,
T. Yanagida and J. Yokoyama Phys. Rev. Lett. {\bf 70}, 1912 (1993); T. Gherghetta and
G. Jungman, Phys. Rev. {\bf D48}, 1546 (1993), A. Acker, H. Kikuchi, E. Ma and
U. Sarkar, Phys. Rev. {\bf D48}, 5006 (1993); P. J. O' Donnell and U. Sarkar,
Phys. Rev. {\bf D49}, 2118 (1994); M. P. Worah, Phys. Rev. {\bf D53}, 3902 (1996);
 R. Jeannerot, Phys. Rev. Lett. {\bf 77}, 3292 (1996); L. Covi, E. Roulet and
F. Vissani, Phys. Lett. {\bf B384}, 169 (1996); W. Buchmuller and M. Plumacher,
Phys. Lett. {\bf B389}, 73 (1996); G. Lazarides, Q. Shafi and N. D. Vlachos,
Phys. Lett. {\bf B427}, 53 (1998); J. Liu and G. Segre, Phys. Rev. {\bf D48},
4609 (1993); M. Flanz, E. A. Paschos, U. Sarkar and J. Weiss, Phys. Lett.
{\bf B389}, 693 (1996).

\bibitem{recent}Recent discussions include:M. S. Berger and B. Brahmachari,
Phys. Rev. {\bf D60}, 073009 (2000); D. Falcone and F. Tramantano,
Phys. Rev. {\bf D63}, 073007 (2001); F. Bucella, Phys. Lett. {\bf
B524}, 241 (2002); H. B. Nielsen and Y. Takanishi, Phys. Lett.
{\bf B507}, 241 (2001); M. Hirsch and S. F. King, Phys. Rev. {\bf
D64}, 113005 (2001); Z. Z. Xing, Phys. Lett. {\bf B545}, 352
(2002); P. H. Frampton, S. L. Glashow and T. Yanagida, Phys. Lett.
{\bf B548}, 119 (2002); T. Endoh et al, Phys. Rev. Lett. {\bf 89},
231601 (2002); S. Davidson and A. Ibarra, Nucl. Phys. {\bf B648},
345 (2003); G. C. Branco et al arXiv:hep-ph/0211001; J. Pati,
Phys. Rev. {\bf D68}, 072002 (2003);
 J. Ellis and M. Raidal, arXiv:hep-ph0206174;
W. Buchmuller, P. Di Bari and M. Plumacher, arXiv:/0205349;
 S. Pascoli, S. T. Petcov and W. Rodejohann; arXiv:hep-ph/0302054;
 E. Kh. Akhmedov, M. Frigerio and A. Yu. Smirnov, arXiv:/hep-ph0305322;
V. Barger, D. A. Dicus, H-J. He and T. Li, Phys. Lett.
{\bf B583}, 173 (2004).

\bibitem{pil}See A. Pilaftsis in \cite{reviews} above.

\bibitem{jpr}A. S. Joshipura, E. A. Paschos and W. Rodejohann,
J. H. E. P. {\bf 0108}, 029 (2001); W. Rodejohann, Phys. Lett.
{\bf B452}, 100 (2002).

\bibitem{rge}See for examples, J. A. Casas et al, Nucl. Phys.
{\bf B573}, 652 (2000) and S. Autusch et al, Nucl. Phys.
{\bf B674}, 401 (2003).

\bibitem{bdp}A recent discussion is given by W. Buchmuller, P. Di Bari
and M. Plumacher in \cite{recent} above.


\bibitem{kt}See Kolb and Turner in \cite{reviews} above.

\bibitem{washout}We use Fig 5a in \cite{bdp} to roughly estimate
the $D_i$ when necessary. Note that the horizontal axis is
interpreted as roughly $10^{-3}K_i$ and the vertical axis as
roughly $2 \times 10^{-8}D_i$.

\bibitem{so10studies}
K. Matsuda, Y. Koide, T. Fukuyama and H. Nishiura, Phys. Rev. {\bf
D65}, 033008 (2002), E 079904; B. Bajc, G. Senjanovic and F.
Vissani, Phys. Rev. Lett. {\bf 90}, 051802 (2003); K. Babu and C.
Macesanu, private communication.

\bibitem{jrs}R. Johnson, S. Ranfone and J. Schechter, Phys. Rev.
{\bf 35}, 282 (1987). Here, $Tr(M_\nu)$=0 was studied in an SO(10)
model together with the assumption of a special form for all the
mass matrices. The more recent discovery of a top quark mass
greater than about 90 GeV has ruled out that special form which,
combined with $Tr(M_\nu)$=0, would make $M_H$ a singular matrix; H.
S. Goh, R. N. Mohapatra and S. P. Ng; Phys. Lett {\bf B570}, 215
(2003).

\end{thebibliography}
\end{document}